\documentclass[twocolumn]{aastex631} 

\newcommand{\numax}{$\nu_{\rm max}$}
\newcommand{\dnu}{$\Delta\nu$}
\shorttitle{}
\shortauthors{Tayar \& Joyce}

\usepackage{listings}

\begin{document}

\title{Star-crossed Clusters: Asteroseismic Ages for Individual Stars are in Tension with\\the Ages of their Host Clusters}

\author[0000-0002-4818-7885]{Jamie Tayar}
\email{jtayar@ufl.edu}
\affiliation{Department of Astronomy, University of Florida, Bryant Space Science Center, Stadium Road, Gainesville, FL 32611, USA}

\author[0000-0002-8717-127X]{Meridith Joyce}
\email{mjoyce8@uwyo.edu}
\affiliation{Department of Physics and Astronomy, University of Wyoming, 1000 E University Ave, Laramie, WY 82071, USA}
\affiliation{School of Computing, University of Wyoming, 1000 E University Ave, Laramie, WY 82071, USA}

\begin{abstract}
A meta-analysis of seismic ages determined for individual stars in the well-studied open and globular clusters NGC 6819, NGC 6791, M67, M4, M19, M80, and M9 reveals both high variance across measurements and significant discrepancy with independent, isochrone-based age determinations for the clusters in which these stars reside. The scatter among asteroseismic ages for individual stars in any one of these clusters far surpasses both the absolute age uncertainty computed for reference cluster M92 (5.4\%) and the model-to-model systematic uncertainties in isochrones (roughly 10\%). This suggests that either binary processes are significantly altering the masses of stars in these clusters, or some additional corrections, perhaps as a function of mass, metallicity, or surface gravity, are required to bring the asteroseismic age scale into concordance with ages inferred from isochrone or similar model fitting. 
\end{abstract}

\section{ Introduction} 
\label{introduction}
The ages of stars are the ``holy grail'' of stellar demographics, 
enabling the study of the formation and evolution of our Galaxy. Open and globular star clusters have long served as the gold standard for stellar age determinations. Originally theorized to be groups of single stars born at the same time and formed from the same gas but spanning a range of birth masses, clusters provided ideal age calibration environments. Theoretical models of single-aged, chemically homogeneous populations, known as isochrones, are relatively straightforward to generate by interpolating over single-star stellar evolutionary tracks with different masses (e.g.\ \citealt{Dotter2016}). 
However, closer study of star clusters and better understanding of stellar multiplicity have revealed additional complexities. We now know that globular clusters may have significant second populations with a range of chemical compositions and possibly slightly 
different ages \citep{BastianLardo2018}. Open clusters may show variations in abundance within the same population, though these are typically small \citep{Sinha2024}. 
Younger clusters also show spreads on the main sequence that could represent an age dispersion \citep[e.g.][]{MackeyBrobyNielsen2007}, but these are more commonly attributed to a distribution of rotation rates, with an upper limit on the age dispersion of tens of millions of years \citep[e.g.][]{Lipatov2022}. 
To coerce cluster data into a more idealized form suitable for model comparison, photometric \citep[e.g.][]{Milone2017a} and spectroscopic \citep[e.g.][]{Schiavon2024} information can be used to separate out members with anomalous chemistry, and it is sometimes possible to identify and isolate binary stars \citep{GodoyRivera2021b}. Similarly, binary interaction products \citep{MathieuGeller2009,Leiner2019} can be removed to provide a more appropriate single star sequence for estimating the cluster age using isochrones.

It is now well-understood that globular and open clusters are not perfect systems with homogeneous populations, but they nonetheless remain the best objects for age determinations. By carefully collating and cleaning observational data, it is possible to use the morphology of clusters on color--magnitude diagrams (CMDs) to infer their ages via model fitting (e.g.\ \citealt{Vandenberg1990, Salaris1998, MarinFranch2009,Dotter2011, VandenBerg2013}).
By quantifying the impacts of variations in the assumptions made about the physics of stellar interiors adopted in the models, it is even possible to calculate age uncertainties from such fits \citep{Ying2023,Ying2024, Reyes2024}. Accounting likewise for uncertainties in distances, composition, nuclear reaction rates, convection physics, and so on, the best cluster ages have been shown to have absolute age uncertainties as low as 5.5\% \citep{Ying2023}.

While clusters provide robust age benchmarks, the majority of stars we observe today do not reside in clusters, thus necessitating the development of a variety of other techniques to estimate stellar ages. However, most of these techniques are still fundamentally calibrated using cluster data.
It is possible in some cases, for example, to use isochrones to derive ages for individual stars (non-cluster members) that lie in especially sensitive regions of the HRD, e.g., sub-giants near or just past the main sequence turn-off \citep{TangJoyce2021, GodoyRivera2021a, Joyce2023}.
However, the reliability of isochrones in general is assessed according to their ability to fit clusters, and so the models themselves adopt clusters implicitly as ground-truth \citep[e.g.][]{Choi2018b}.

Another age determination technique is gyrochronology, which maps the rotation periods and temperatures of stars to their ages using the fact that low-mass dwarf stars spin down over time \citep{Barnes2003, Angus2015, Bouma2024}. This technique is empirically calibrated to provide absolute rather than relative ages using rotating stars in open clusters \citep{Rebull2018, Curtis2020}. Chemical ratios may also be correlated with stellar ages, either because of the principles of galactic chemical evolution (e.g.\ [Y/Mg], \citealt{daSilva2012,Berger2022}) or due to internal mixing, which is traced by diagnostics such as Li \citep{Martin2018, GalindoGuil2022} and [C/N] \citep{Martig2016, Roberts2024}. Such diagnostics must be calibrated either directly on clusters \citep{Spoo2022} or on intermediate results \citep{Pinsonneault2018} that are themselves calibrated to clusters. 
%

%
Another technique that has been gaining popularity for age inference on galactic scales is asteroseismology. Asteroseismology uses the global oscillation properties of a star to estimate its stellar mass \citep{Brown1991, KjeldsenBedding1995}. From there, it is possible to estimate an age that is significantly less sensitive to assumptions about the internal stellar physics than using, for example, a luminosity and temperature (L. Morales et al., in prep.). While there are variations of asteroseismology that try to reproduce the frequencies of individual oscillation modes to estimate extremely precise ages for targets with high-fidelity signals \citep{SilvaAguirre2017,LiT2024, LiY2024, Joyce2024}, such techniques are computationally expensive and not amenable to large numbers of stars. 
We focus instead on methods that can be applied to samples of red giants large enough to perform galactic archaeology. 
In these techniques, the observed large frequency spacing (separation between adjacent $p$-modes), $\Delta \nu$, and frequency of maximum power, $\nu_{\text{max}}$, are combined with information about the temperature and metallicity to estimate a stellar mass, radius, and age \citep[e.g.][]{Schonhutstasik2024, Pinsonneault2024}. However, studies using binaries \citep{Gaulme2016}, open clusters \citep{Pinsonneault2018}, Gaia \citep{GaiaDR3, Zinn2019}, and models \citep{White2011, Mosser2013, Sharma2016} have suggested that asteroseismic inferences made using simple scaling relations may not be accurate,
and corrections have been developed to improve the estimated radii and masses.

It is therefore prudent to ask how asteroseismic ages compare to cluster-based age determinations. In this analysis, we compare the ages estimated from asteroseismic analyses of individual, first-ascent red giants in well-studied open and globular clusters to the ages determined for the clusters themselves, by independent means. We describe the collation of the heterogeneous asteroseismic data from individual cluster analyses, and compare the inferred ages to the cluster ages derived from isochrone fitting. We show the scale of the uncertainties from the individual scatter within clusters as well as the cluster age scale. Finally, we argue that in order to ensure that the tens to hundreds of thousands of field star ages expected from CoRoT \citep{Anders2017}, Kepler \citep{Pinsonneault2024}, K2 \citep{Stokholm2023, Warfield2024}, TESS \citep{Hon2021,TheodoridisTayar2023}, Roman (N. Downing, in prep.), and PLATO \citep{Miglio2017} are on a true and accurate scale, reanalysis of existing data and the collection of additional cluster data are required. 

\begin{deluxetable*}{rrrrrrr}
\label{Tab:isoages}
\tablecaption{Assumed cluster ages from isochrone fitting, assumed cluster metallicities, the source of the asteroseismology and whether or not first-ascent red giants with good seismic constraints are available, as used in this work. The metallicities reported in this table are mainly iron abundances [Fe/H], but some are metal abundances [M/H], measurements often dominated by iron lines. We do not report $\alpha$-enhancements for every cluster in our table although some are $\alpha$-element enhanced (see text).  Note that M9 is plotted at 13.2\,Gyr in the figures for ease of viewing.   }
\tablehead{\colhead{Cluster} & \colhead{Age (Gyr)} & \colhead{Age Source} &\colhead {[Fe/H]} &\colhead{[Fe/H] Source} &\colhead{Seismic Source} &\colhead{RGB} }
\startdata
NGC 6819 & 2.54 & \cite{Sanjayan2022} & 0.03 & \cite{Myers2022} & \cite{Pinsonneault2024} & Y \\ 
M67 & 3.95 & \cite{Reyes2024} & 0 & \cite{Myers2022} & \cite{Stello2016} & Y \\ 
NGC 6791 & 8.3 & \cite{Brogaard2012} & 0.31 & \cite{Myers2022} & \cite{Pinsonneault2024} & Y \\ 
M4 & 11.5 & \cite{VandenBerg2013} & $-1.1$ & \cite{Howell2022} & \cite{Howell2022} & Y \\ 
M19 & 12 & \cite{Howell2025} & $-1.55$ & \cite{Howell2025} & \cite{Howell2025} & Y \\ 
M80 & 13 & \cite{Howell2024} & $-1.6$ & \cite{Howell2024} & \cite{Howell2024} & Y \\ 
M9 & 13 & \cite{Howell2025} & $-1.791$ & \cite{Howell2025} & \cite{Howell2025} & Y \\ 
NGC 6633 &	0.45&	\cite{Smiljanic2009}&	$-0.03$ &	\cite{Lagarde2015}	&\cite{Lagarde2015}&	N\\
NGC 6866 & 0.65 & \cite{CantatGaudin2020} & 0 & \cite{Myers2022} & \cite{Brogaard2023} & N \\ 
Hyades & 0.775 & \cite{Brandner2023} & 0.18 & \cite{Brandner2023} & \cite{Arentoft2019} & N \\ 
NGC 6811 & 1 & \cite{Sandquist2016} & $-0.06$ & \cite{Myers2022} & \cite{Sandquist2016} & N \\ 
NGC 1817 & 1 & \cite{Sandquist2020} & $-0.16$ & \cite{Myers2022} & \cite{Sandquist2020} & N \\
NGC 2506 & 2.01 & \cite{Knudstrup2020} & $-0.36$ & \cite{Knudstrup2020} & \cite{Knudstrup2020} & N \\ 
\enddata

\end{deluxetable*}

\section{ Known Uncertainties}
While data do exist that could enable homogeneous cluster age determinations on a broader scale, the current literature often quotes a wide range of disparate and discrepant ages for the same cluster. Because ages are an inherently model-dependent inference, some variance in inferred cluster ages has been correctly attributed to instrumental uncertainties, modeling systematics, or the adoption of different physical assumptions within models. 
That said, the dispersions between quoted ages can be quite large, suggesting that additional effort to bring clusters onto a homogeneous scale may still be required. For example, for the well-studied open cluster M67, recent work cites ages from 3.8\,Gyr \citep{Nataf2024} to 4.3\,Gyr \citep{CantatGaudin2020}, and in NGC 6791, ages ranging from 6\,Gyr \citep{AnthonyTwarogTwarog1985} to 12\,Gyr \citep{Stetson2003} have been claimed. 
The metallicities quoted for a given cluster in the literature can likewise show dispersion that is often larger than the uncertainties quoted in any particular paper \citep[see e.g.][for a more thorough indication of the diversity of ages and metallicities in the literature for any individual cluster]{Wu2014b}. We can therefore say, in general, that the age inferred depends on both the model used and the method of inference \citep{ByromTayar2024}. That said, while a complete and homogeneous reanalysis of the cluster age scale across the full range of clusters \citep{CantatGaudin2020, HuntReffert2024} is a worthy goal, it represents a significant effort outside the scope of this paper. Instead, we use in this analysis a representative age for each cluster from the recent literature, acknowledging that while scatter exists, we do not expect any individual choice to affect our result. Because we understand the true uncertainty in the cluster ages to be dominated by method-to-method scatter, we also do not propagate the individual error bars of any particular study.

Similarly, different seismic analyses of the same star will often report systematically different results for \numax\ and $\Delta \nu$ \citep{Pinsonneault2018,Pinsonneault2024}. Solar values adopted for \numax, $\Delta \nu$, and $T_{\text{eff}}$ differ across analyses \citep{Pinsonneault2018}, and there are likewise
different corrections \citep{White2011,Mosser2013, Sharma2016} and different external calibrations that can be applied \citep{Zinn2019,Pinsonneault2018,Pinsonneault2024}. Similar to the cluster age case, the method-to-method differences in choices can have a much larger impact on the final value than the quoted random uncertainties from the individual measurements, and so we do not quote the individual uncertainties reported internally for each asteroseismic target here. In the best cases, multiple methods of analysis, including comparison between several different asteroseismic pipelines, are done on the same stars, a variety of different corrections and calibration schemes are implemented, and the uncertainties should rightfully include both the systematic differences from the methods as well as the true observational uncertainties \citep{Pinsonneault2024}. That said, while doing such an analysis for the combined cluster sample would be extremely valuable as a future endeavor, it is a significant undertaking that is outside of the scope of this paper.

\section{ Methods}
\label{methods}
To compare the asteroseismic age scale to the age scale from clusters, we collected data from the literature on seismic detections of first-ascent red giant branch stars in well-characterized open and globular clusters. The ages and other stellar parameters for each of these stars along with their host cluster are presented in Table \ref{Tab:seismoages}.
While there are oscillation detections from other types of pulsators in clusters \citep{Bedding2023,Molnar2024}, we restrict our analysis to the solar-like oscillators on the first-ascent red giant branch that are the most relevant for age calibration of galactic archaeology studies. This choice limits this work to the 7 clusters listed in the first seven rows of Table \ref{Tab:isoages}.

For M67, we assumed a cluster (i.e.\ isochrone-based) age of 3.95\,Gyr \citep{Reyes2024} and obtained seismic ages for individual red giant members from \cite{Stello2016}, although other results do exist \citep[e.g.][]{LiT2024,JoChang2024}. %
For NGC 6791 and NGC 6819, we used the most recent seismic ages as published in \citet{Pinsonneault2024} and membership from the OCCAM DR17 catalog \citep{Myers2022}, requiring a membership probability of greater than 0.3 from proper motion, radial velocity, metallicity, and \citet{CantatGaudin2020}, although again, other analyses of seismic samples do exist in the literature \citep[e.g.][]{Basu2011, Hekker2011,Stello2011a, Stello2011b, Corsaro2012, Wu2014a, Handberg2016, Corsaro2017, Handberg2017, McKeever2019, Brogaard2021, Cakirli2022, CorveloPaz2023, JoChang2024}. Ages for these clusters were assumed to be 8.3\,Gyr \citep{Brogaard2012} and 2.54\,Gyr \citep{Sanjayan2022}, respectively.
We note that while stars from NGC 6811 are also in this field and thus in the catalog, consistent with previous works \citep[e.g.][]{Arentoft2017, Sandquist2016}, there were no seismic detections in first-ascent red giants that were also high-probability cluster members and so we do not include them. 

For stars that did not already have published ages, we used a grid of models optimized for the estimation of ages from first-ascent red giants published in \citet{Tayar2017} using the masses, surface gravities, metallicities, and $\alpha$-element abundances as inputs.
This method of estimating ages is in particular not sensitive to the temperature and therefore the mixing length assumed in the modeling \citep{JoyceTayar2023} and only sensitive at roughly the 10\% level to model-to-model systematics in general (L. Morales et al., in prep.). 
For M4, seismic ages were determined using masses, surface gravities, and temperatures presented in \citet{Howell2022}, along with their stated metallicity of $[\mathrm{Fe}/\mathrm{H}] = -1.1$ and $[\alpha/\mathrm{Fe}] =0.4$, for each star. We note that seismic results for some stars are also available \citep{Miglio2016,Tailo2022, JoChang2024}.  The cluster age was assumed to be 11.5\,Gyr \citep{VandenBerg2013}.
For M80, we assumed a cluster age of 13\,Gyr, and masses, surface gravities, and temperatures as published in \citet{Howell2024}, assuming their stated metallicity of $[\mathrm{Fe}/\mathrm{H}]  = -1.791$ and $[\alpha/\mathrm{Fe}] =0.4$ for each star. 
For M9 and M19 \citep{Howell2025}, we used the same procedure to turn the masses, surface gravities, assumed cluster metallicities of 
$-1.67$ and $-1.55$, respectively, and $\alpha$-element enhancements of $+0.4$ for both clusters into ages for all of the stars listed as RGB in \citet{Howell2025}. These were compared to cluster ages of 13\,Gyr and 12\,Gyr, respectively, quoted in that work. For ease of visualization on the figures, we show M9 at an age of 13.2\,Gyr.

For NGC 1817, we examined two stars identified as possible red giants, adopting \numax\ and \dnu\ from \citet{Sandquist2020}; one of these stars was also identified by \citet{JoChang2024}. We assumed the cluster metallicity of $-0.1$ as stated in the \citet{Sandquist2020} paper, and used the seismic scaling relations with solar values of $\nu_{\rm max, \odot}=3090\,\mu\mathrm{Hz}$,  $\delta\nu_{\odot}= 135.1 \,\mu\mathrm{Hz}$, and $\mathrm{T}_{\rm eff, \odot}=5771.8\,\mathrm{K}$ \citep{Huber2011} to compute masses for these stars. However, the resulting ages were offset from the 1\,Gyr age assumed for the cluster by roughly 200\%. Assuming that either the evolutionary state identification for these stars is incorrect, or more careful work is needed, we exclude them from our figure. 

We note that seismic measurements are also available for stars in NGC 6811 \citep[e.g.][]{Hekker2011,Stello2011a, Stello2011b,Corsaro2012,Sandquist2016, Corsaro2017, Arentoft2017, JoChang2024}, the Hyades \citep{Lund2016,Arentoft2019}, NGC 6866 \citep{Brogaard2023}, NGC 6633 \citep{Poretti2015,Lagarde2015}, and NGC 2506 \citep{Knudstrup2020}. However, these clusters are young, and all of their stars are either not yet on the giant branch or have been reported being in the core helium-burning (CHeB) phase rather than as first-ascent red giants. This means that inferring ages requires careful treatment of advanced evolutionary stages, including and especially mass loss, that is beyond the scope of this paper.

\section{ Age Comparison}
\begin{figure*}
\includegraphics[width=\textwidth]{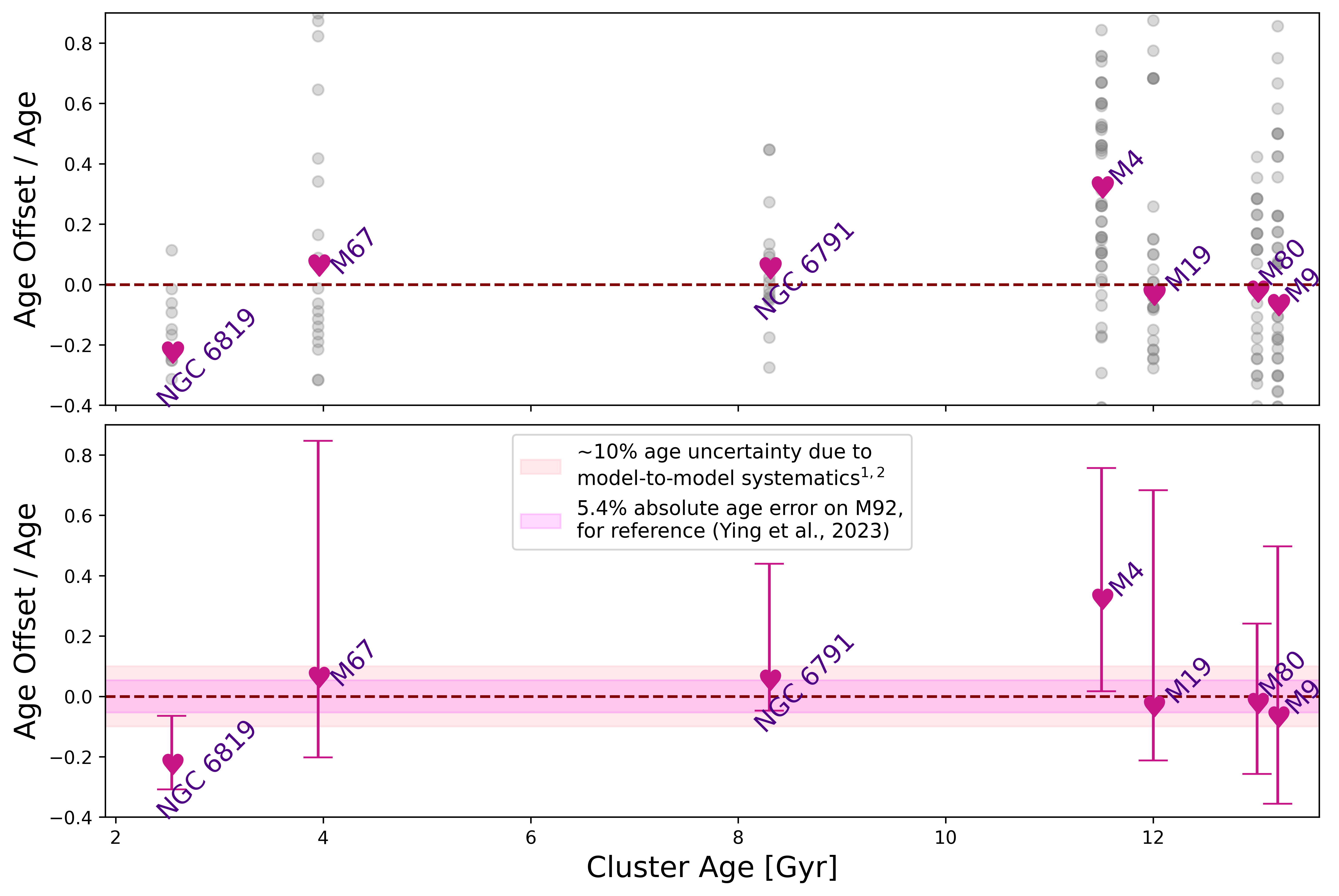}
\caption{ For seven well-characterized open and globular clusters, asteroseismic ages for individual members are compared to cluster ages derived from isochrone fitting. The colored bands represent the expected uncertainties on both the cluster age derivation (pink) and the asteroseismic age determination \citep[orange, ][]{Joyce2023,Pinsonneault2024} in the optimal case. We find that the scatter for individual stars within the same cluster (grey circles) far exceeds these expected uncertainties (top). Additionally, even if we average over all the red giants in a cluster (bottom) there may be systematic offsets as a function of age which may require future calibration. The data for M9 are artificially shifted to appear at 13.2\,Gyr rather than the literature age of 13\,Gyr for ease of visualization. }
\label{fig:clusters}
\end{figure*}

In Figure \ref{fig:clusters}, we compare the ages inferred from the asteroseismic analysis of each individual red giant, independent of its cluster membership, to the age inferred for the whole cluster from isochrone fitting. In the upper panel, asteroseismic age determinations for individual stars are shown as grey circles in terms of \textit{age offset}, computed via [seismic age $-$ cluster age] / cluster age, versus cluster age. In both panels, the median value among asteroseismic ages for each cluster is shown as a pink heart, as a function of cluster age. 
In the lower panel, asymmetric dispersions of the asteroseismic ages are shown instead of individual points. These represent the 16th and 84th quantiles and serve as a rough measure of variance among heterogeneous measurements for any particular cluster.

Notably, for some clusters, such as M67 and NGC 6791, the median age offsets are consistent within the uncertainties intrinsic to isochrone-based determinations of cluster ages \citep[estimated to be at least 6\% according to calculations for the comparable cluster M92 from][]{Ying2023}, indicated via the dark pink horizontal shaded region, or the model-to-model scatter of ages inferred this way \citep[$\sim$10\%,][L. Morales et al., in prep.]{Pinsonneault2024}, indicated in light orange. 
However, for a number of the clusters, the systematic offsets between the two ages significantly exceed these uncertainties (e.g.\  NGC 6819, M4). 

More strikingly, the dispersions among asteroseismic ages within each cluster dramatically eclipse both the model-to-model systematic uncertainties and the best-case scenario absolute age error. In all clusters but NGC 6819, asteroseismic dispersion exceeds the model-to-model systematic uncertainties by a factor of at least~5 (and a factor of 10 in the case of M67). 
Because of the inhomogeneous nature of the asteroseismic analysis and age determination methods, it is possible that some of the offsets result from the particular method of seismic determination \citep{Pinsonneault2018}, the grid of models used \citep{ByromTayar2024}, or the heterogeneous determination of cluster ages. That said, it still appears that there may be underlying systematic offsets between the cluster and asteroseismic ages scales.
To determine whether such a systematic offset between the age scales exists will require isolating and removing the corrupting effects of differences in asteroseismic analysis techniques as well as a careful reinvestigation of the data within each cluster. Systematic, homogeneous asteroseismic age determinations, ideally with the same observatory and analysis pipeline, would be a worthwhile endeavor.

\begin{figure*}
\includegraphics[width=\textwidth]{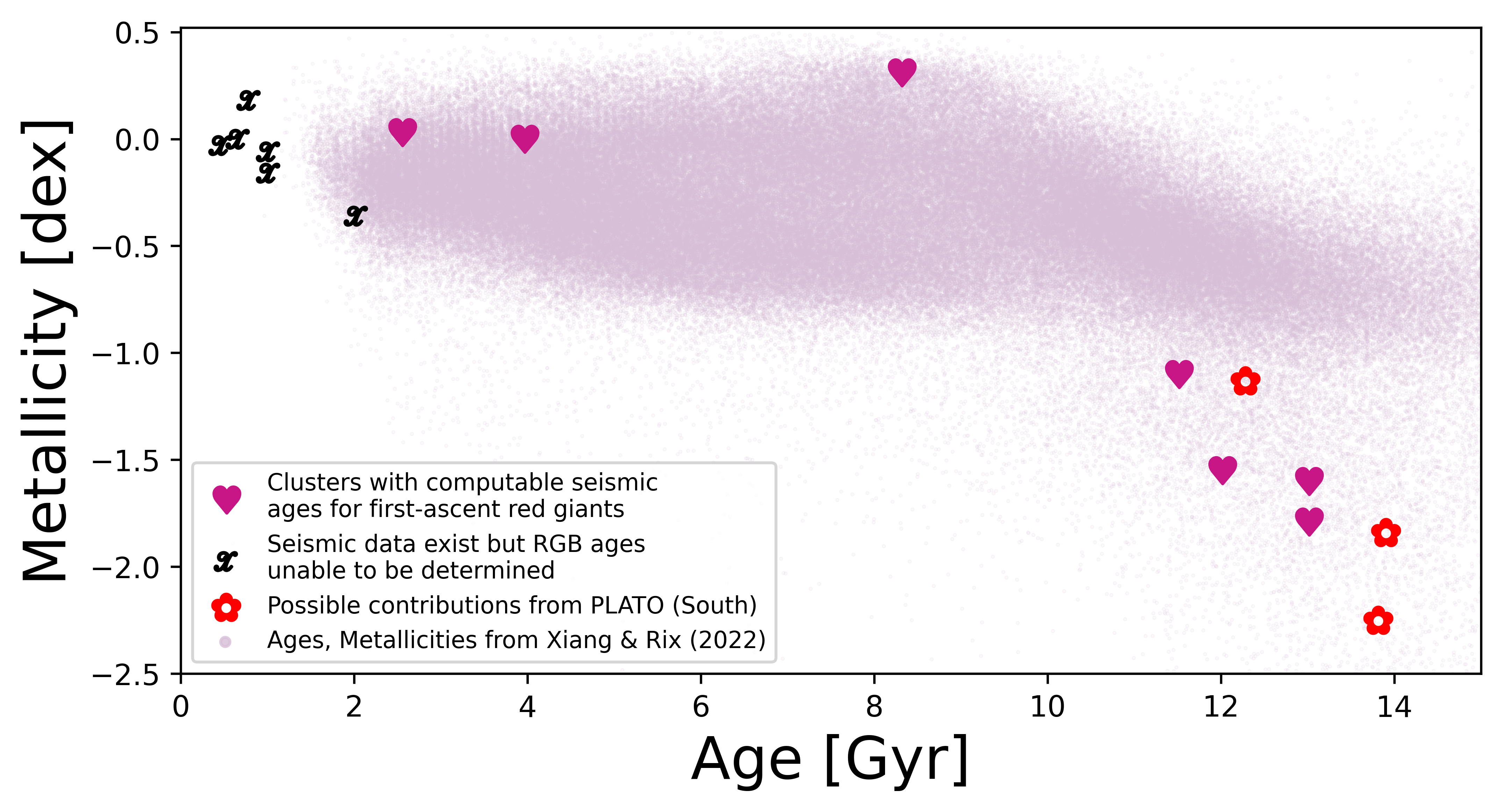}
\caption{ We show the distribution of ages for a large sample of Milky Way stars \citep[small dots, ][] {XiangRix2022} as a function of age and metallicity. The clusters that we discuss here, which can act as calibrators for the asteroseismic age estimates, are shown as pink hearts. Clusters for which any seismic data exist, even if the data is of clump stars rather than first-ascent giants, or for which we were not able to determine reliable ages, are shown as black Xs. We also mark the location of potential clusters in the upcoming PLATO mission (red flowers, see text). However, we find that these cluster populations are not well distributed among the ages and metallicities of the majority of stars in the Milky Way, posing a challenge for the detailed calibration of the asteroseismic age scale.   
}
\label{fig:galactic}
\end{figure*}

\section{ Galactic Context}
Asteroseismic ages for red giants are being used to study the properties of stars across the Milky Way and dissect the history of the Galaxy itself \citep{SilvaAguirre2018,Stokholm2023}, as well as serving as calibrators for other age techniques including machine learning and chemical diagnostics \citep[e.g.][]{Martig2016, Ness2016, Leung2023, StoneMartinez2024}. However, as demonstrated in the previous section, there is reason to doubt that those asteroseismic ages have been properly calibrated to the more fundamental cluster age scale. 

We also show in Figure \ref{fig:galactic} that the majority of clusters with asteroseismic data---those that could currently be used to calibrate such results---are not particularly well-matched to the distribution of galactic ages and metallicities inferred from other methods \citep[e.g.\ isochrone fitting of subgiants,][]{XiangRix2022}. When compared to the clusters shown in this analysis (pink hearts) and all other clusters for which any seismic data exist (black markers), there are still very few constraints in the most populous parts of the age--metallicity parameter space.  
Figure \ref{fig:galactic} also demonstrates that future missions \citep[e.g.\ PLATO,][]{PLATO} capable of adding new clusters with asteroseismic red giant detections to this valuable sample would provide a significant advancement to the field in this context. Using the predictions from \citet{Nascimbeni2022}, we show in Figure \ref{fig:galactic} clusters for which we anticipate receiving reliable red giant oscillation detections using ages \citep{Ying2023, Valcin2020} and metallicities \citep{Schiavon2024} from the literature. While these plans should improve the density of calibrators at old ages, there are still many regions of this parameter space devoid of calibrators.

\section{ Future Work}
Asteroseismic ages are becoming available for large numbers of red giant stars at high precision. However, ages sit at the end of long chains of inference in stellar parameter determinations
and often have systematic offsets that are challenging to constrain. Most methods for age estimation ultimately address this concern by calibrating their results to clusters, which provide the best constraints on the properties of stars of different masses. Asteroseismic data for red giants in clusters have started to become available at a range of ages and metallicities, but our initial meta-analysis of these results suggests that additional work on this topic is required. 
We uncovered significant dispersions among inferred ages of stars in the same cluster, as well as systematic shifts between the average cluster age inferred from asteroseismology, and the benchmark values from isochrone fitting available in the literature. While this has been noted before in individual clusters, we show here that it is a feature broadly present across current data.  

Careful analysis of a small number of high-value targets has the potential to disentangle the cluster age scale versus asteroseismic age scale discrepancy. Systematically derived, seismic ages for just a handful of stars in well-constrained clusters whose ages are independently determined using isochrone fitting would provide a critical calibration baseline that is currently absent from the literature, and while missions have been proposed to carry out such a study \citep[e.g.\ HAYDN,][]{HAYDN}, none have yet flown. 
In much the same way that asteroseismologists have developed corrections to the \numax\ scale \citep{Pinsonneault2018} and 
\dnu\ scaling relation \citep{White2011}, we anticipate that asteroseismologists will determine a correction to the age scale.
Development of such a correction would allow the placement of the hundreds of thousands of asteroseismic ages that are becoming available \citep[][A. Theodoridis et al., in prep.]{Stokholm2023,Warfield2024, Pinsonneault2024} from missions like CoRoT, Kepler, K2, TESS, and soon PLATO onto an accurate scale that can be used for galactic archaeology. Such studies have already begun to provide interesting insight into the relative evolution of the Milky Way disks \citep{SilvaAguirre2018}, and the history of Milky Way interactions \citep{Chaplin2020, Grunblatt2021}. Placing huge numbers of stars on a reliable absolute scale will enable us to constrain the evolutionary history of the Milky Way and unlock new understanding about the star formation, chemical enrichment, and merger history of our galactic home.

\begin{acknowledgements}
J.\ Tayar and M.\ Joyce contributed equally to this manuscript and may both refer to this as a first-author publication.
J.T. would like to thank the UF Astraeus Space Institute for financial support for this effort. J.T. would also like to thank Open Cluster Journal Club and Russel White for helpful cluster thoughts, and Susie Byrom for her M67 literature review. 
J.T. and M.J. wish to thank Sarah Ballard, Alex Camazon, Zach Claytor, Isabel Colman, Jason Dittman, Anthony Gonzalez, Rafael Guzman, Corin Marasco, and Joel Zinn for helpful discussions about cluster seismology. 
\end{acknowledgements}

\bibliography{clusters.bib}{}
\bibliographystyle{aasjournal}


\startlongtable
\begin{deluxetable}{rrrrrrrr}
\label{Tab:seismoages}
\tablecaption{Ages inferred for individual cluster first-ascent red giants. For each star, we have agnostically reported information as discussed in the source papers and interpolated ages in the \citet{Tayar2017} grid if they were not reported in the original work. ID numbers are KIC IDs or TIC IDs where stated in the original papers. If the paper only referred to cluster members by numbers, we have appended a number related to the cluster name and several zeros before that number, to avoid conflicting notation. Because of the heterogeneous nature of these sources, and the desire to understand the scatter between clusters rather than the individual uncertainties of each target, we do not report uncertainties on each column here, although some may be provided in the original source. }
\tablehead{\colhead{ID} & \colhead{Mass} & \colhead{Logg} & \colhead{[M/H]} & \colhead{[$\alpha$/M]} & \colhead{Teff} & \colhead{Cluster} & \colhead{Age}}  
\startdata
246884141 & 1.36 & 2.46 &  $-$0.1 & 0 & 5109 & NGC1817 & 3.69 \\ 
246893162 & 1.29 & 2.35 &  $-$0.1 & 0 & 4787 & NGC1817 & 4.46 \\ 
8000009 & 0.59 & 1.4 &  $-$1.791 & 0.4 & 4700 & M80 &  $-$99 \\ 
80000101 & 1.49 & 1.8 &  $-$1.791 & 0.4 & 4554 & M80 & 1.61 \\ 
80000103 & 0.89 & 1.88 &  $-$1.791 & 0.4 & 4705 & M80 & 9.07 \\ 
80000105 & 0.89 & 1.8 &  $-$1.791 & 0.4 & 4605 & M80 & 9.08 \\ 
80000107 & 0.85 & 1.45 &  $-$1.791 & 0.4 & 4454 & M80 & 10.69 \\ 
80000110 & 0.86 & 1.8 &  $-$1.791 & 0.4 & 4670 & M80 & 10.24 \\ 
80000113 & 0.93 & 1.97 &  $-$1.791 & 0.4 & 4746 & M80 & 7.76 \\ 
80000117 & 0.84 & 1.72 &  $-$1.791 & 0.4 & 4654 & M80 & 11.14 \\ 
80000120 & 0.79 & 1.91 &  $-$1.791 & 0.4 & 4739 & M80 & 13.90 \\ 
80000123 & 0.81 & 1.95 &  $-$1.791 & 0.4 & 4756 & M80 & 12.71 \\ 
80000124 & 0.81 & 1.83 &  $-$1.791 & 0.4 & 4691 & M80 & 12.72 \\ 
80000127 & 0.83 & 1.66 &  $-$1.791 & 0.4 & 4597 & M80 & 11.64 \\ 
80000128 & 0.76 & 1.69 &  $-$1.791 & 0.4 & 4666 & M80 & 15.96 \\ 
80000129 & 0.77 & 1.75 &  $-$1.791 & 0.4 & 4689 & M80 & 15.23 \\ 
80000132 & 0.78 & 1.66 &  $-$1.791 & 0.4 & 4635 & M80 & 14.55 \\ 
80000141 & 0.9 & 1.87 &  $-$1.791 & 0.4 & 4721 & M80 & 8.73 \\ 
80000190 & 0.77 & 2.1 &  $-$1.791 & 0.4 & 4834 & M80 & 15.21 \\ 
80000194 & 0.81 & 2 &  $-$1.791 & 0.4 & 4717 & M80 & 12.71 \\ 
80000199 & 0.75 & 2.13 &  $-$1.791 & 0.4 & 4846 & M80 & 16.72 \\ 
80000203 & 0.81 & 1.97 &  $-$1.791 & 0.4 & 4701 & M80 & 12.71 \\ 
80000220 & 0.78 & 1.97 &  $-$1.791 & 0.4 & 4750 & M80 & 14.53 \\ 
80000222 & 0.87 & 2.2 &  $-$1.791 & 0.4 & 4823 & M80 & 9.80 \\ 
80000234 & 0.87 & 2.02 &  $-$1.791 & 0.4 & 4811 & M80 & 9.81 \\ 
80000236 & 0.78 & 2.09 &  $-$1.791 & 0.4 & 4874 & M80 & 14.52 \\ 
8000041 & 0.81 & 1.83 &  $-$1.791 & 0.4 & 4728 & M80 & 12.72 \\ 
8000045 & 0.75 & 1.7 &  $-$1.791 & 0.4 & 4656 & M80 & 16.75 \\ 
8000046 & 0.77 & 1.73 &  $-$1.791 & 0.4 & 4643 & M80 & 15.23 \\ 
8000050 & 0.75 & 1.75 &  $-$1.791 & 0.4 & 4670 & M80 & 16.75 \\ 
8000056 & 0.82 & 1.54 &  $-$1.791 & 0.4 & 4532 & M80 & 12.17 \\ 
8000058 & 0.73 & 1.57 &  $-$1.791 & 0.4 & 4584 & M80 & 18.51 \\ 
8000068 & 0.76 & 1.49 &  $-$1.791 & 0.4 & 4567 & M80 & 15.97 \\ 
8000069 & 0.74 & 1.88 &  $-$1.791 & 0.4 & 4774 & M80 & 17.58 \\ 
4000010 & 0.86 & 2.60 &  $-$1.1 & 0.4 & 4960 & M4 & 12.17 \\ 
40000104 & 0.79 & 2.44 &  $-$1.1 & 0.4 & 4851 & M4 & 16.81 \\ 
4000011 & 0.71 & 1.93 &  $-$1.1 & 0.4 & 4612 & M4 & 24.74 \\ 
40000124 & 0.79 & 2.88 &  $-$1.1 & 0.4 & 5088 & M4 & 16.70 \\ 
40000129 & 0.85 & 2.76 &  $-$1.1 & 0.4 & 4989 & M4 & 12.68 \\ 
4000013 & 0.79 & 2.60 &  $-$1.1 & 0.4 & 4840 & M4 & 16.83 \\ 
40000130 & 0.85 & 2.48 &  $-$1.1 & 0.4 & 4897 & M4 & 12.72 \\ 
40000131 & 0.84 & 2.90 &  $-$1.1 & 0.4 & 5047 & M4 & 13.20 \\ 
40000133 & 0.74 & 1.41 &  $-$1.1 & 0.4 & 4377 & M4 & 21.22 \\ 
4000014 & 0.84 & 2.43 &  $-$1.1 & 0.4 & 4807 & M4 & 13.29 \\ 
40000142 & 0.89 & 2.78 &  $-$1.1 & 0.4 & 4988 & M4 & 10.74 \\ 
40000143 & 0.86 & 2.73 &  $-$1.1 & 0.4 & 4937 & M4 & 12.15 \\ 
40000147 & 0.88 & 2.94 &  $-$1.1 & 0.4 & 5218 & M4 & 11.14 \\ 
40000148 & 0.66 & 1.99 &  $-$1.1 & 0.4 & 4632 & M4 & 32.55 \\ 
4000015 & 0.68 & 2.16 &  $-$1.1 & 0.4 & 4865 & M4 & 28.96 \\ 
4000016 & 0.77 & 2.56 &  $-$1.1 & 0.4 & 4921 & M4 & 18.35 \\ 
40000165 & 0.92 & 2.89 &  $-$1.1 & 0.4 & 5287 & M4 & 9.49 \\ 
40000169 & 0.96 & 2.83 &  $-$1.1 & 0.4 & 5143 & M4 & 8.13 \\ 
4000017 & 0.92 & 2.62 &  $-$1.1 & 0.4 & 4785 & M4 & 9.54 \\ 
40000180 & 0.75 & 2.32 &  $-$1.1 & 0.4 & 4972 & M4 & 20.05 \\ 
40000188 & 0.83 & 2.40 &  $-$1.1 & 0.4 & 4810 & M4 & 13.89 \\ 
40000190 & 0.78 & 2.70 &  $-$1.1 & 0.4 & 4949 & M4 & 17.54 \\ 
40000191 & 0.77 & 2.52 &  $-$1.1 & 0.4 & 4892 & M4 & 18.36 \\ 
40000194 & 0.76 & 2.82 &  $-$1.1 & 0.4 & 5035 & M4 & 19.16 \\ 
40000195 & 0.61 & 1.84 &  $-$1.1 & 0.4 & 4687 & M4 & 43.23 \\ 
40000197 & 0.81 & 2.14 &  $-$1.1 & 0.4 & 4734 & M4 & 15.24 \\ 
4000020 & 0.85 & 1.94 &  $-$1.1 & 0.4 & 4535 & M4 & 12.78 \\ 
40000205 & 0.82 & 2.71 &  $-$1.1 & 0.4 & 4971 & M4 & 14.48 \\ 
4000021 & 0.84 & 2.42 &  $-$1.1 & 0.4 & 4855 & M4 & 13.29 \\ 
40000210 & 0.78 & 2.93 &  $-$1.1 & 0.4 & 5093 & M4 & 17.44 \\ 
40000213 & 0.82 & 2.64 &  $-$1.1 & 0.4 & 4836 & M4 & 14.51 \\ 
40000215 & 0.91 & 3.03 &  $-$1.1 & 0.4 & 5174 & M4 & 9.85 \\ 
40000216 & 0.84 & 2.38 &  $-$1.1 & 0.4 & 4813 & M4 & 13.29 \\ 
40000217 & 1.01 & 2.63 &  $-$1.1 & 0.4 & 5015 & M4 & 6.80 \\ 
4000022 & 0.75 & 2.46 &  $-$1.1 & 0.4 & 4852 & M4 & 20.19 \\ 
40000225 & 0.72 & 1.42 &  $-$1.1 & 0.4 & 4427 & M4 & 23.43 \\ 
40000229 & 0.8 & 2.11 &  $-$1.1 & 0.4 & 4677 & M4 &  $-$99 \\ 
4000023 & 0.76 & 2.54 &  $-$1.1 & 0.4 & 4897 & M4 & 19.23 \\ 
40000238 & 0.87 & 2.84 &  $-$1.1 & 0.4 & 5065 & M4 & 11.63 \\ 
4000024 & 0.87 & 2.56 &  $-$1.1 & 0.4 & 4829 & M4 & 11.69 \\ 
40000244 & 0.82 & 2.42 &  $-$1.1 & 0.4 & 4877 & M4 & 14.52 \\ 
4000025 & 0.78 & 2.20 &  $-$1.1 & 0.4 & 4719 & M4 & 17.61 \\ 
40000252 & 0.78 & 2.66 &  $-$1.1 & 0.4 & 5012 & M4 & 17.49 \\ 
4000026 & 0.7 & 1.67 &  $-$1.1 & 0.4 & 4541 & M4 & 26.04 \\ 
4000027 & 0.8 & 2.76 &  $-$1.1 & 0.4 & 4954 & M4 &  $-$99 \\ 
4000028 & 0.79 & 3.02 &  $-$1.1 & 0.4 & 5157 & M4 & 16.55 \\ 
4000029 & 0.79 & 2.69 &  $-$1.1 & 0.4 & 4986 & M4 & 16.76 \\ 
4000030 & 0.76 & 2.40 &  $-$1.1 & 0.4 & 4913 & M4 & 19.19 \\ 
4000031 & 0.81 & 2.67 &  $-$1.1 & 0.4 & 4966 & M4 & 15.17 \\ 
4000032 & 0.81 & 2.80 &  $-$1.1 & 0.4 & 5026 & M4 & 15.14 \\ 
4000034 & 0.79 & 2.98 &  $-$1.1 & 0.4 & 5059 & M4 & 16.59 \\ 
4000035 & 0.83 & 2.43 &  $-$1.1 & 0.4 & 4838 & M4 & 13.89 \\ 
40000358 & 0.77 & 2.34 &  $-$1.1 & 0.4 & 4814 & M4 & 18.38 \\ 
4000036 & 1.01 & 2.45 &  $-$1.1 & 0.4 & 4808 & M4 & 6.82 \\ 
40000363 & 0.71 & 1.80 &  $-$1.1 & 0.4 & 4633 & M4 & 24.67 \\ 
40000365 & 0.75 & 1.71 &  $-$1.1 & 0.4 & 4540 & M4 & 20.20 \\ 
4000037 & 0.85 & 2.46 &  $-$1.1 & 0.4 & 4821 & M4 & 12.72 \\ 
40000408 & 0.77 & 2.05 &  $-$1.1 & 0.4 & 4672 & M4 & 18.42 \\ 
4000048 & 0.82 & 2.16 &  $-$1.1 & 0.4 & 4746 & M4 & 14.56 \\ 
190000161 & 0.78 & 1.52 &  $-$1.55 & 0.4 & 4732 & M19 & 15.12 \\ 
190000162 & 0.84 & 1.55 &  $-$1.55 & 0.4 & 4610 & M19 & 11.58 \\ 
190000183 & 0.97 & 1.54 &  $-$1.55 & 0.4 & 4497 & M19 & 6.95 \\ 
190000198 & 0.7 & 1.63 &  $-$1.55 & 0.4 & 4734 & M19 & 22.49 \\ 
190000242 & 0.83 & 1.74 &  $-$1.55 & 0.4 & 4769 & M19 & 12.09 \\ 
190000247 & 0.8 & 1.84 &  $-$1.55 & 0.4 & 4897 & M19 & 13.83 \\ 
190000258 & 0.83 & 1.84 &  $-$1.55 & 0.4 & 4950 & M19 & 12.08 \\ 
190000261 & 0.84 & 1.89 &  $-$1.55 & 0.4 & 4876 & M19 & 11.57 \\ 
190000271 & 0.85 & 1.9 &  $-$1.55 & 0.4 & 4874 & M19 & 11.08 \\ 
190000276 & 0.72 & 1.81 &  $-$1.55 & 0.4 & 4891 & M19 & 20.21 \\ 
190000288 & 0.84 & 1.8 &  $-$1.55 & 0.4 & 4904 & M19 & 11.57 \\ 
190000315 & 0.89 & 1.95 &  $-$1.55 & 0.4 & 4844 & M19 & 9.41 \\ 
190000331 & 0.9 & 1.89 &  $-$1.55 & 0.4 & 4718 & M19 & 9.06 \\ 
190000335 & 0.85 & 1.93 &  $-$1.55 & 0.4 & 4829 & M19 & 11.08 \\ 
190000336 & 0.81 & 2.05 &  $-$1.55 & 0.4 & 4955 & M19 & 13.21 \\ 
190000352 & 0.85 & 2.06 &  $-$1.55 & 0.4 & 4940 & M19 & 11.07 \\ 
190000355 & 0.9 & 2.19 &  $-$1.55 & 0.4 & 4895 & M19 & 9.04 \\ 
190000357 & 0.85 & 2.43 &  $-$1.55 & 0.4 & 5112 & M19 & 11.04 \\ 
190000359 & 0.81 & 2.14 &  $-$1.55 & 0.4 & 4994 & M19 & 13.20 \\ 
190000360 & 0.71 & 2.19 &  $-$1.55 & 0.4 & 5039 & M19 & 21.28 \\ 
190000365 & 0.66 & 2.13 &  $-$1.55 & 0.4 & 4924 & M19 & 28.04 \\ 
190000366 & 0.67 & 2.33 &  $-$1.55 & 0.4 & 5095 & M19 & 26.44 \\ 
190000367 & 0.72 & 2.24 &  $-$1.55 & 0.4 & 5124 & M19 & 20.17 \\ 
190000368 & 0.72 & 2.2 &  $-$1.55 & 0.4 & 5155 & M19 & 20.16 \\ 
190000389 & 0.87 & 2.39 &  $-$1.55 & 0.4 & 5058 & M19 & 10.16 \\ 
190000392 & 0.89 & 2.13 &  $-$1.55 & 0.4 & 4918 & M19 & 9.40 \\ 
190000394 & 0.88 & 2.3 &  $-$1.55 & 0.4 & 4975 & M19 & 9.77 \\ 
190000395 & 0.84 & 2.27 &  $-$1.55 & 0.4 & 4970 & M19 & 11.54 \\ 
190000396 & 0.72 & 2.08 &  $-$1.55 & 0.4 & 4934 & M19 & 20.22 \\ 
190000399 & 0.8 & 2.24 &  $-$1.55 & 0.4 & 4949 & M19 & 13.82 \\ 
190000401 & 0.91 & 2.37 &  $-$1.55 & 0.4 & 5035 & M19 & 8.67 \\ 
190000403 & 0.84 & 2.24 &  $-$1.55 & 0.4 & 5011 & M19 & 11.54 \\ 
190000422 & 0.82 & 2.09 &  $-$1.55 & 0.4 & 5113 & M19 & 12.61 \\ 
90000103 & 0.71 & 1.4 &  $-$1.67 & 0.4 & 4618 & M9 & 20.87 \\ 
90000122 & 0.7 & 1.73 &  $-$1.67 & 0.4 & 4784 & M9 & 22.00 \\ 
90000130 & 0.94 & 1.61 &  $-$1.67 & 0.4 & 4635 & M9 & 7.61 \\ 
90000136 & 0.68 & 1.49 &  $-$1.67 & 0.4 & 4678 & M9 & 24.53 \\ 
90000137 & 0.91 & 1.77 &  $-$1.67 & 0.4 & 4601 & M9 & 8.54 \\ 
90000164 & 0.98 & 2.24 &  $-$1.67 & 0.4 & 4828 & M9 & 6.54 \\ 
90000185 & 0.79 & 2.06 &  $-$1.67 & 0.4 & 4870 & M9 & 14.13 \\ 
90000188 & 0.91 & 2.23 &  $-$1.67 & 0.4 & 4849 & M9 & 8.50 \\ 
90000189 & 0.83 & 2.44 &  $-$1.67 & 0.4 & 5110 & M9 & 11.78 \\ 
90000192 & 0.78 & 2.02 &  $-$1.67 & 0.4 & 5033 & M9 & 14.77 \\ 
90000195 & 0.73 & 2.16 &  $-$1.67 & 0.4 & 4979 & M9 & 18.78 \\ 
90000197 & 0.76 & 1.97 &  $-$1.67 & 0.4 & 4826 & M9 & 16.23 \\ 
90000198 & 0.73 & 2.26 &  $-$1.67 & 0.4 & 4865 & M9 & 18.81 \\ 
90000201 & 0.62 & 2.25 &  $-$1.67 & 0.4 & 4988 & M9 & 34.41 \\ 
90000209 & 0.82 & 2.22 &  $-$1.67 & 0.4 & 4844 & M9 & 12.34 \\ 
90000210 & 0.51 & 2.21 &  $-$1.67 & 0.4 & 5146 & M9 &  $-$99 \\ 
90000214 & 0.79 & 2.19 &  $-$1.67 & 0.4 & 4869 & M9 & 14.13 \\ 
90000215 & 0.77 & 2.04 &  $-$1.67 & 0.4 & 4875 & M9 & 15.47 \\ 
90000222 & 0.85 & 2.19 &  $-$1.67 & 0.4 & 4855 & M9 & 10.83 \\ 
90000223 & 0.93 & 2.28 &  $-$1.67 & 0.4 & 4818 & M9 & 7.86 \\ 
90000224 & 0.89 & 2.1 &  $-$1.67 & 0.4 & 4827 & M9 & 9.21 \\ 
9000023 & 0.76 & 1.84 &  $-$1.67 & 0.4 & 4754 & M9 & 16.24 \\ 
9000024 & 0.83 & 1.5 &  $-$1.67 & 0.4 & 4529 & M9 & 11.84 \\ 
90000240 & 0.76 & 2.3 &  $-$1.67 & 0.4 & 4994 & M9 & 16.20 \\ 
90000241 & 0.89 & 2.28 &  $-$1.67 & 0.4 & 4856 & M9 & 9.20 \\ 
90000242 & 0.82 & 2.08 &  $-$1.67 & 0.4 & 4942 & M9 & 12.35 \\ 
90000247 & 0.85 & 2.15 &  $-$1.67 & 0.4 & 4970 & M9 & 10.83 \\ 
90000248 & 0.87 & 2 &  $-$1.67 & 0.4 & 4844 & M9 & 9.98 \\ 
90000249 & 1.02 & 2.16 &  $-$1.67 & 0.4 & 4690 & M9 & 5.70 \\ 
90000250 & 0.69 & 1.97 &  $-$1.67 & 0.4 & 5107 & M9 & 23.13 \\ 
90000258 & 0.78 & 1.93 &  $-$1.67 & 0.4 & 4781 & M9 & 14.79 \\ 
90000259 & 1.04 & 2.37 &  $-$1.67 & 0.4 & 4840 & M9 & 5.31 \\ 
90000264 & 0.93 & 2.31 &  $-$1.67 & 0.4 & 4921 & M9 & 7.86 \\ 
90000281 & 0.87 & 2.06 &  $-$1.67 & 0.4 & 4799 & M9 & 9.97 \\ 
90000287 & 0.77 & 2.07 &  $-$1.67 & 0.4 & 4898 & M9 & 15.47 \\ 
90000289 & 0.79 & 1.93 &  $-$1.67 & 0.4 & 4851 & M9 & 14.14 \\ 
90000292 & 0.86 & 2.15 &  $-$1.67 & 0.4 & 4797 & M9 & 10.39 \\ 
90000295 & 1.13 & 2.3 &  $-$1.67 & 0.4 & 4976 & M9 & 4.01 \\ 
9000035 & 0.72 & 1.57 &  $-$1.67 & 0.4 & 4732 & M9 & 19.80 \\ 
9000041 & 0.72 & 1.84 &  $-$1.67 & 0.4 & 4799 & M9 & 19.80 \\ 
9000054 & 0.85 & 1.58 &  $-$1.67 & 0.4 & 4598 & M9 & 10.86 \\ 
9000066 & 0.74 & 1.94 &  $-$1.67 & 0.4 & 4900 & M9 & 17.87 \\ 
9000067 & 0.89 & 1.82 &  $-$1.67 & 0.4 & 4603 & M9 & 9.23 \\ 
9000075 & 0.72 & 1.86 &  $-$1.67 & 0.4 & 4830 & M9 & 19.80 \\ 
9000081 & 0.54 & 1.64 &  $-$1.67 & 0.4 & 4900 & M9 &  $-$99 \\ 
9000085 & 0.84 & 1.67 &  $-$1.67 & 0.4 & 4581 & M9 & 11.33 \\ 
9000088 & 0.79 & 1.41 &  $-$1.67 & 0.4 & 4532 & M9 & 14.16 \\ 
2436209 & 1.18 & 2.65 & 0.324 & 0.012 & 4424 & NGC6791 & 7.86 \\ 
2436332 & 1.14 & 2.34 & 0.357 & 0.041 & 4298 & NGC6791 & 9.05 \\ 
2569204 & 1.27 & 1.62 & 0.281 & 0.025 & 3926 & NGC6791 & 6.02 \\ 
2436540 & 1.14 & 2.66 & 0.309 & 0.007 & 4421 & NGC6791 & 8.75 \\ 
2569360 & 1.15 & 2.22 & 0.270 & 0.023 & 4260 & NGC6791 & 8.49 \\ 
2436688 & 1.09 & 2.77 & 0.316 & 0.024 & 4471 & NGC6791 & 10.56 \\ 
2436814 & 1.05 & 2.28 & 0.291 & 0.038 & 4299 & NGC6791 & 12.01 \\ 
2436884 & 1.17 & 1.83 & 0.294 & 0.025 & 4021 & NGC6791 & 8.05 \\ 
2436900 & 1.13 & 2.44 & 0.257 & 0.044 & 4363 & NGC6791 & 8.84 \\ 
2569618 & 1.17 & 2.64 & 0.300 & 0.036 & 4427 & NGC6791 & 8.17 \\ 
2437340 & 1.17 & 1.80 & 0.298 & 0.032 & 4020 & NGC6791 & 7.94 \\ 
2569935 & 1.04 & 1.60 & 0.320 & 0.035 & 4006 & NGC6791 & 12.01 \\ 
2437496 & 1.11 & 1.52 & 0.289 & 0.030 & 3898 & NGC6791 & 9.41 \\ 
2437507 & 1.23 & 2.20 & 0.308 & 0.033 & 4243 & NGC6791 & 6.84 \\ 
2437653 & 1.18 & 2.77 & 0.319 & 0.056 & 4534 & NGC6791 & 7.91 \\ 
2437965 & 0.84 & 1.74 & 0.311 & 0.029 & 4107 & NGC6791 & 25.81 \\ 
2570518 & 1.13 & 2.56 & 0.284 & 0.054 & 4456 & NGC6791 & 9.15 \\ 
2438333 & 1.14 & 2.68 & 0.319 & 0.074 & 4504 & NGC6791 & 8.74 \\ 
2708270 & 1.15 & 2.03 & 0.287 & 0.035 & 4165 & NGC6791 & 8.37 \\ 
2571093 & 0.93 & 1.16 & 0.300 & 0.025 & 3724 & NGC6791 & 17.66 \\ 
5111718 & 1.65 & 3.04 & 0.101 & 0.002 & 4901 & NGC6819 & 2.30 \\ 
5023732 & 1.60 & 2.34 &  $-$0.004 & 0.012 & 4578 & NGC6819 & 2.38 \\ 
5111940 & 1.59 & 2.63 & 0.043 & 0.007 & 4726 & NGC6819 & 2.50 \\ 
5112072 & 1.65 & 3.01 & 0.039 &  $-$0.025 & 4886 & NGC6819 & 2.16 \\ 
5024272 & 3.49 & 2.16 & 0.073 &  $-$0.008 & 4661 & NGC6819 & 0.26 \\ 
5024297 & 1.74 & 2.57 & 0.044 & 0.003 & 4696 & NGC6819 & 1.90 \\ 
5024583 & 1.78 & 2.48 & 0.034 &  $-$0.006 & 4652 & NGC6819 & 1.74 \\ 
5112734 & 1.73 & 2.51 & 0.057 & 0.002 & 4641 & NGC6819 & 1.93 \\ 
5112744 & 1.72 & 2.55 & 0.017 & 0.006 & 4676 & NGC6819 & 1.95 \\ 
5112880 & 1.98 & 2.31 & 0.027 &  $-$0.008 & 4596 & NGC6819 & 1.26 \\ 
5112948 & 1.71 & 2.53 & 0.045 & 0.003 & 4713 & NGC6819 & 1.98 \\ 
5113061 & 1.53 & 1.54 & 0.031 & 0.005 & 4183 & NGC6819 & 2.83 \\ 
4937576 & 1.72 & 2.42 & 0.004 & 0.017 & 4616 & NGC6819 & 1.90 \\ 
5113441 & 1.69 & 3.10 & 0.078 & 0.019 & 4885 & NGC6819 & 2.12 \\ 
\enddata

\end{deluxetable}


\end{document}